\documentclass[a4 paper, 12 pt] {article}

\begin{document}
\title{\bf{Non-Abelian Color Dielectric \\-- towards the Effective Model
of the Low Energy QCD}}
\author{A. Wereszczy\'{n}ski $^{a)}$ \thanks{wereszcz@alphas.if.uj.edu.pl} \, and
{M. \'{S}lusarczyk $^{a,b)}$ \thanks{mslus@phys.ualberta.ca}}
       \\
       \\ $^{a)}$ Institute of Physics,  Jagiellonian University,
       \\ Reymonta 4, Krak\'{o}w, Poland
       \\
       \\  $^{b)}$ Department of Physics, University of Alberta,
       \\ Edmonton, Alberta T6G 2J1, Canada}
\maketitle
\begin{abstract}
Lattice motivated triplet color scalar field theory is analyzed.
We consider non-minimal as well as covariant derivative coupling
with $SU(2)$ gauge fields. Field configurations generated by
external electric sources are presented. Moreover non-Abelian
magnetic monopoles are found. Dependence on the spatial
coordinates in the obtained solutions is identical as in the usual
Abelian case. We show also that after a decomposition of the
fields a modified Faddeev--Niemi action can be obtained. It
contains explicit $O(3)$ symmetry breaking term parameterized by
the condensate of an isoscalar field. Due to that Goldstone bosons
observed in the original Faddeev--Niemi model are removed.
\end{abstract}
\newpage
\section{\bf{ Introduction}}
At present there exist several approaches to the dynamics of
non-Abelian gauge fields in infrared region. The most popular one
-- lattice QCD (see e.g. \cite{creutz}) -- deals with the theory
on the most fundamental level and leads to many important results
concerning hadron spectrum. However, an unavoidable ingredient of
this class of methods are numerical studies which are in fact
their main drawback. If the method itself was correct and relevant
numerical simulations produced output comparable with an
experiment, the most fascinating dynamics of gluonic fields would
be still veiled in numerics. The situation is even worse since the
results of ongoing lattice studies remain controversial.

The whole bunch of phenomenological models in continuum
space--time has been proposed as an alternative to lattice gauge
theory. To call only a few let us mention here stochastic vacuum
model \cite{dosch},  various realizations of the Abelian
projection and monopole dominance (see e.g. \cite{shabanov}), the
Faddeev--Niemi model \cite{niemi1, niemi2}, and color dielectric
models. All these models attempt to describe a number of  known
features of non-perturbative QCD but their relation to the
original theory is usually not completely clear. In particular
none of them has been strictly derived from the underlying
fundamental theory so far.

A promising step in this direction was done by Nielsen and Patkos
a long time ago \cite{nielsen}. They pointed out that color
dielectric models can be derived from the lattice QCD by the
so-called blocking procedure described in more detail in the next
section. In principle this method can be used to obtain color
dielectric model from the full theory as a result of strict
well--defined sequence of steps. However, the procedure turned out
to be far too complicated to be accomplished in practice. Every
realization of this scenario known to date either leads only to
partial results or is based on strong and probably not well
justified assumptions.

The color dielectric models derived from lattice theory have
certain common features which do not depend on details of the
blocking procedure \cite{mathiot}. In particular field contents of
the resulting effective model is universal for all these
approaches. It allows to construct an effective action with
requirements of its invariance with respects to Lorentz and color
group instead of attempting strict derivation. As we show in this
paper this set of small and natural assumptions astonishingly
restricts the family of possible models.

In the discussion presented below we consider a color dielectric
model with non-Abelian gauge fields and non-Abelian scalar
dielectric field. We show that form of the action we deal with is
a natural consequence of chosen set of field degrees of freedom
and invariance requirements. Our model easily reproduces
confinement of external sources known from commonly discussed
models with ordinary scalar field. In addition, after redefining
degrees of freedom by mens of the generalized Faddeev--Niemi--Cho
decomposition we end up with  modified Faddeev--Niemi action which
can be applied do describe physical excitations in the glueball
sector. In  contradiction to the original Faddeev--Niemi model our
action explicitly brakes $O(3)$ invariance. To the best of our
knowledge this is the first time when this fundamental property is
studied in color dielectric type model.

The plan of our paper is the following. In the next section we
define relevant degrees of freedom and postulate simple Lorentz
and gauge invariant action. Then we solve field equations for the
model and demonstrate how the confining potential for external
sources as well as magnetic monopoles emerge in the framework of
the discussed theory. In section \ref{fnsec} we perform
decomposition of fields and obtain an effective theory of
Faddeev--Niemi type corresponding to our non-Abelian color
dielectric model in the glueball sector. Finally we briefly
summarize our results and propose some possible extensions of the
present discussion.
\section{\bf{ Non-Abelian color dielectric action}}
It is well known that pure $SU(N)$ gauge theory can be defined on
the lattice in terms of unitary link variables $U_k(x)$, where $x$
denotes beginning of the link and $k=1, \dots ,4$ its orientation.
The standard Wilson action has the form of products of these link
matrices summed over plaquettes (see e.g. \cite{creutz}) and in
the continuum limit corresponds to the normal Yang--Mills
invariant. The Wilson formulation has been successfully applied to
many numerical simulations on discrete lattice. However, due to
large quantum fluctuations of the original gluonic degrees of
freedom continuum limit in non-perturbative regime is cumbersome.

In a phenomenological model proposed a long time ago by Friedberg
and Lee \cite{cdiel1} the confinement is modeled by a scalar field
coupled non-minimally to the non-Abelian gauge field, which can be
treated as a low--momentum component of the original gluon field.
The Friedberg--Lee model constituted a new way of thinking about
non-perturbative dynamics of gluon fields and was interesting on
its own but it was also too arbitrary to be seriously considered
as physically well justified theory of strong interactions in
certain momenta region. However, the gap between Wilson approach
and color dielectric models has been bridged by Nielsen and Patkos
\cite{nielsen} and later by Mack \cite{mack} who showed that
effective action of the kind of that by Friedberg and Lee can be
derived from lattice formulation. This scenario was then explored
and developed by Pirner and collaborators who studied a few
versions of color dielectric models derived from lattice
\cite{arodz, dalley}.

The common feature of all these attempts was introduction of
averaged, infrared stable collective variables instead of
fluctuating gluon fields. In the Pirner's approach original
$SU(3)$ fields $U_k(x)$ were averaged over larger space regions
and eventually formed more general $3 \times 3$ matrix variable
$M_k (x)$ defined on coarse--grained lattice.

In general, any $N\times N$ matrix can be decomposed in the
following way
\begin{equation}
M_k (x)=\hat{V}_k (x) \hat{\chi}_k (x) e^{i\theta_k(x)}
\label{decomposition}
\end{equation}
where $\hat{V}_k$ is a unitary $SU(N)$ matrix, $\hat{\chi}_k$ is a
positively defined hermitian $N \times N$ matrix and $\theta_k$ is
a real number. One can observe that $N=2$ case is special. Then
$\theta_k = 0$ and $\hat{\chi}$ becomes a positive number. The
interpretation of these field is still a little bit mysterious.
Usually one relates $\hat{V}_k$ to a traceless hermitian gauge
field $\hat{A}_k$ - lattice gluon field:
$$\hat{V}_k(x)=e^{i\hat{A}_k(x)}.$$ Analogously, $\theta_k$ is a new Abelian
gauge field. The last field $\hat{\chi_k}$ is know as a color
dielectric field.

Transformation properties of the introduced lattice field can be
simply deducted from their definition \cite{mathiot}. $\theta_k$
turns out to have standard vector transformation law as expected.
On the contrary the color dielectric field transforms as:
\begin{equation}
\hat{\chi}_k(x) = \hat{\chi}_{-k}(x + b e_k),
\end{equation}
where $b$ denotes lattice spacing and therefore it cannot be
regarded as a Lorentz vector. This problem has been addressed many
times (see e.g. \cite{mathiot, arodz}) but no ultimate conclusion
has been made so far. In the discussion below we restrict
ourselves to the simplest possibility taking $\hat{\chi_k}$ as a
scalar.

In principle the form of color dielectric action should strictly follow
from the macroscopic theory by the blocking procedure. Unfortunately,
this point of the approach remains spurious and no commonly accepted color
dielectric action has been directly derived yet. Many authors
agree that the relevant physical features are given by the diagonal part of the
color dielectric field. Thus, one takes:
\begin{equation}
\hat{\chi}_k(x)=\chi_k(x) \, I \label{diel_variable1}
\end{equation}
where $I$ is $N \times N$ unit matrix. However, there are no clear
arguments that this approximation is valid i.e. that the
off-diagonal degrees of freedom do not influence the physics of
the model. Due to that one should investigate the full non-Abelian
color dielectric field. We follow \cite{arodz} and introduce a new
color dielectric field
\begin{equation}
\hat{\chi}_k(x)=(\hat{\phi}_k(x))^2, \label{arodz_variable}
\end{equation}
where
\begin{equation}
\hat{\phi}_k(x)=\phi_k(x) \, I + \lambda_a \phi^a_k(x).
\label{diel_variable2}
\end{equation}
Here $\lambda^a$ are Gell-Mann matrices in the case of $SU(3)$
group.
\\
In this paper we propose a classical continuum effective action
based on the gauge fields and color dielectric field.
This means that we replace lattice variables, defined in the four
dimensional Euclidean space-time, by continuous fields in
Minkowski space-time
\begin{equation}
\{ \hat{A}_k, \, \theta_k, \, \hat{\phi}_k \} \longrightarrow \{
\hat{A}_{\mu}, \, \theta_{\mu}, \, \hat{\phi} \}
\label{limit_cont}
\end{equation}
One should be aware that there are a lot of questions concerning
this substitution. However, as we are here mainly interested in
the definition of the relevant variables for the low energy
gluodynamics the exact form of the continuous limit is not so
important (it inflects the action not the fields). Due to that we
treat this problem in the most naive way. Moreover, we assume
'minimal non-trivial' situation that is we neglect Abelian gauge
field $\theta_{\mu}$. In addition the full color gauge group
$SU(3)$ is substituted by its subgroup $SU(2)$. It is worth noting
that starting the whole construction from $2 \times 2$ matrices
leads us to the standard scalar (in Lorentz as well as in color
group) color dielectric field and not to $SU(2)$ field. However,
we are aiming at studying the simplest imaginable model with
non-Abelian dielectric function as a introductory step to the
proper gauge group even if it does not fit the general pattern.
Thus we take
\begin{equation}
\hat{\phi }=\sigma_a \phi^a, \label{diel_variable3}
\end{equation}
where $\sigma^a$ are Pauli matrices, $a=1,2,3$. In the realistic
situation one should obviously deal with the full gauge group. As
one can easily check adding the diagonal part $\phi I$ in
(\ref{diel_variable3}) does not change the results obtained below.
\\
Let us now proceed to construction of the action for our model. As
it was mentioned above we do not attempt to derive it strictly
from the lattice formulation. Instead, we build the simplest
Lorentz and gauge invariant action using previously chosen fields.

The Abelian color dielectric field couples usually to the gauge
field via the so-called dielectric function. Here situation is
more subtle. On account of the fact that color dielectric field is
a vector in the color space it transforms in the fundamental
representation under the gauge transformation. Thus the standard
derivative in the kinetic term has to be replaced by the covariant
one $D_{\mu }\phi^a$. It is the most natural way of preserving
gauge invariance of the gradient term. On the other hand color
dielectric mechanism demands non-minimal coupling of dielectric
function $\sigma$ to the standard Yang--Mills invariant. This is
the crucial point -- the coupling between color dielectric and
gauge field is double folded. As we show it in the next sections
the minimal coupling is connected with glueball (topological)
sector of the model whereas non-minimal coupling assures
confinement of external electric sources.
\\
To conclude, the simplest non-Abelian gauge and Lorentz invariant
extension of the color dielectric action reads as follows
\begin{equation}
S = \int d^{4}x \left[ -\frac{1}{4} \sigma \left( \frac{ \phi_a
\phi^a }{\Lambda^2 } \right) F^{a}_{\mu \nu} F^{a \mu \nu} +
\frac{1}{2} (D_{\mu} \phi^a) (D^{\mu} \phi_a) \right],
\label{action}
\end{equation}
where the covariant derivative is defined in the standard manner
\begin{equation}
D_{\mu } \phi^a = \partial_{\mu } \phi^a - \epsilon^{abc} A_{\mu
}^b \phi^c. \label{cov}
\end{equation}
and dielectric function is assumed to have usual form, known from
color dielectric models \cite{dick2, my1}:
\begin{equation}
\sigma = \left( \frac{\phi_a \phi^a}{\Lambda^2} \right)^{4\delta
}, \label{dick}
\end{equation}
where $\delta >\frac{1}{4}$ and $\Lambda $ is a dimensional
constant setting an energy scale in the model. For completeness,
let us mention a dielectric function which allows for the linear
confinement \cite{dick1}
\begin{equation}
\sigma = \exp \left( b \frac{ \phi_a \phi^a }{\Lambda^2} \right).
\label{dilaton}
\end{equation}
Other dielectric functions has been also considered (see. e.g.
\cite{chabab, bazeia}). In general, such model can include also a
potential term for the non-Abelian color dielectric field. Then
the vacuum value of this field is fixed by the minimum of the
potential. In our investigation, for simplicity reasons, the
potential will be dropped and asymptotic value of $\vec{\phi}$ is
a free parameter.
\\
The pertinent equations of motion for the model defined above read
\begin{equation}
D_{\mu } \left[ \sigma \left( \frac{ \phi_a \phi^a }{\Lambda^2 }
\right) F^{a \mu \nu} \right] = \epsilon^{abc} \phi^b D_{\nu}
\phi^c \label{eqmot1}
\end{equation}
and
\begin{equation}
D_{\mu }D^{\mu } \phi^a + \frac{1}{2} F^{a}_{\mu \nu} F^{a \mu
\nu} \sigma' \phi^a =0, \label{eqmot2}
\end{equation}
where prime denotes differentiation with respect to $\phi^a
\phi_a$. In the next section some solutions of these equations
will be presented.
\section{\bf{ Solutions}}
\subsection{ \bf{Electric case}}
In this subsection a solution with external electric charge will
be constructed. Unfortunately, exact solutions are known only in
the Abelian sector of our model. In the other words, we are forced
to investigate the standard Abelian color dielectric theory.
However, as we it will be shown below even in the Abelian version
our model reproduces confinement of quarks. It seems to be an
advantage of the model that one does not have to deal with the
full non-Abelian theory to find confining solutions. Of course the
role of the non-Abelian degrees of freedom in the confinement
mechanism remains an important issue for further consideration in
the future.
\\
Let us now briefly present the way which leads to the confinement
of an external charge in the color dielectric approach. The
dielectric scalar field was primary introduced to change the long
range behavior of the electric field. The scalar field is needed
to cancel (or weaken) the singularity of the electric field at the
point where the charge is located. The total energy is still
infinite but now it is caused by the behavior of the gauge field
at the spatial infinity. Using this approach one could model
confinement of quarks and get a reasonable inter--quark potential.
\\
The Abelian part of the full color dielectric model can be
obtained by the following restriction:
\begin{equation}
A^a_{\mu} = A_{\mu} \delta^{a3} \label{restriction1}
\end{equation}
and
\begin{equation}
\phi^a =\phi \delta^{a3}. \label{restriction2}
\end{equation}
We consider an external static, point-like electric source:
\begin{equation}
j^a_{\mu } = 4\pi q \delta (r) \delta_{0 \mu} \delta^{a3},
\label{current}
\end{equation}
located at the origin. We assume spherical symmetry of the
problem and introduce the following, purely electric Ansatz
\begin{equation}
\phi= \phi(r) \label{anzatz_el1}
\end{equation}
and
\begin{equation}
E^a_i(r)=-\partial_i U (r) \delta^{a3}, \; \; A^a_i=0,
\label{anzatz_el2}
\end{equation}
where $\phi (r)$ and $U (r)$ are unknown functions. Moreover,
using assumed spherical symmetry one can write $E_i (r)= E(r)
e_i$, where $e_i$ is a unit vector in the $i$ direction.
\\
With this assumptions the field equations can be rewritten in the following
form
\begin{equation}
\left[ r^2 \sigma \left( \frac{ \phi^2}{\Lambda^2} \right) E
\right]'=4\pi q \delta (r) \label{eqmot_el1}
\end{equation}
and
\begin{equation}
\nabla^2_r \phi=-\frac{1}{2} \sigma'_{\phi} E^2_a.
\label{eqmot_el2}
\end{equation}
The first equation can be immediately solved and it gives  a relation
between the scalar and the electric field
\begin{equation}
E(r)=\frac{q}{r^2} \frac{1}{\sigma }. \label{relation}
\end{equation}
Here the role  of the dielectric field is clearly visible.
One can write the last relation as $E(r) =
\frac{q_{eff}}{r^2}$, where $q_{eff}=\frac{q}{\sigma}$ is a new
effective coupling constant. The dielectric field acts now as
a medium in which the 'normal', electric field propagates.
\\
Substituting (\ref{relation}) into (\ref{eqmot_el2}) we derive
differential equation for the scalar field
\begin{equation}
\nabla^2_r \phi=-\frac{q^2}{2r^4} \frac{\sigma'_{\phi}
}{\sigma^2}, \label{eqmot_el_h1}
\end{equation}
which can be integrated for any dielectric function. The general
solution, given by an integral, reads
\begin{equation}
\int \frac{d\phi}{\sqrt{\frac{1}{\sigma (\phi)} +C}} = \frac{1}{r}
+ D, \label{sol_el_gen1}
\end{equation}
where $C$ and $D$ are integration constants. For the dielectric
function (\ref{dick}) proposed previously we obtain the whole
family of solutions label by a positive parameter $\beta_0$
\begin{equation}
E^a_i =A^{-8\delta }  q \frac{x^i}{r^3} \left( \frac{1}{r\Lambda}
+ \frac{1}{\beta_0} \right)^{\frac{-8\delta }{1+4\delta }}
\delta^{a3} \label{sol_el1}
\end{equation}
and
\begin{equation}
\phi^a =\delta^{a3} A \Lambda \left( \frac{1}{r\Lambda} +
\frac{1}{\beta_0} \right)^{\frac{1}{1+4\delta }}, \label{sol_el2}
\end{equation}
$A=[ q(1+4\delta)]^{\frac{1}{1+4\delta}}$. The new constant
$\beta_0$ corresponds to $D=\frac{1}{\beta_0}$. Moreover, because
of the fact that we are looking for finite energy solutions, the
second integration constant $C$ has been set to zero.
\\
Now, electric potential generated by point source is given by
\begin{equation}
U= A^{-8\delta} q \frac{4\delta +1}{4\delta -1} \left(
\frac{1}{r\Lambda} + \frac{1}{\beta_0} \right)^{\frac{1-4\delta
}{1+4\delta }}. \label{sol_el3}
\end{equation}
One can see that for
\begin{equation}
\delta > \frac{1}{4} \label{cond_delta}
\end{equation}
the normal singularity in the electric potential, known from usual
Maxwell theory, no longer exists since  the electric potential
approaches $0$ when $r \rightarrow 0$. One can notice that the
long range behavior of the electric field remains unchanged i.e.
it falls as $\frac{1}{r^2}$. Due to that one can expect that
obtained solutions have finite energy. Precisely speaking, the
corresponding energy density takes the form
\begin{equation}
\varepsilon=A^{-8\delta} \frac{q^2}{r^4} \left( \frac{1}{r\Lambda}
+ \frac{1}{\beta_0} \right)^{\frac{-8\delta }{1+4\delta }}.
\label{endensity_el}
\end{equation}
Then, after integrating over three dimensional space, we find that
the total energy is indeed finite
\begin{equation}
E_N=\Lambda \frac{4\delta +1}{4\delta -1} A^{-8\delta } q^2
\beta_0^{\frac{4\delta -1}{4\delta +1}}. \label{energy_el}
\end{equation}
In addition to presented family of finite energy solutions
there is a singular solution corresponding to $\beta_0 \rightarrow
\infty$:
\begin{equation}
E^a_i = \frac{x^i}{r} A^{-8\delta } q\Lambda^2 \left(
\frac{1}{\Lambda r} \right)^{\frac{2}{1+4\delta }} \delta^{a3}
\label{sol_el_conf1}
\end{equation}
and
\begin{equation}
\phi^a = \delta^{a3} A \Lambda \left( \frac{1}{\Lambda r}
\right)^{\frac{1}{1+4\delta }}. \label{sol_el_conf2}
\end{equation}
In this case the electric potential reads
\begin{equation}
U= qA^{-8\delta} \Lambda \frac{4\delta +1}{4\delta -1} \left(
\frac{1}{\Lambda r} \right)^{\frac{1-4\delta}{1+4\delta }}.
\label{sol_el_conf3}
\end{equation}
This solution describes confining sector of the model. In the
vicinity of the point charge the solution behaves identically as
the finite energy configurations that is the singularity at $r=0$
is removed. However, in contradiction to the previous case, a new
singularity appears. The electric potential diverges at the
spatial infinity  as $r^{\alpha }$ where $\alpha \in (0,1)$. The
standard linear confining potential is reproduced in the limit
when $\delta \rightarrow \infty$. One can easily show that the
same effects are observed if one analyzes the corresponding total
energy.
\\
Such confining inter--quarks potentials, weaker than linear, are
commonly discussed in the framework of non-relativistic potential
models. They have been found in fits to charmonium and bottomium
spectra (see for example Zalewski--Motyka \cite{kacper} and Martin
\cite{martin} potentials). On the other hand, there are some
theoretical arguments based in general on the analytical approach
to QCD, which suggest that energy stored in a flux-tube spanned
between quarks grows slower than linearly \cite{nesterenko}.
\subsection{\bf{ Magnetic case}}
Let us now turn to the purely magnetic sector of the theory. We take
advantage of the well--known spherical magnetic Ansatz:
\begin{equation}
\phi^a = \Lambda \frac{x^a}{r} h(r) \label{anzatz_mag1}
\end{equation}
and
\begin{equation}
A^a_i = \epsilon^{aik} \frac{x^k}{r^2} (g(r)-1),
\label{anzatz_mag2}
\end{equation}
where functions $h$ and $g$ are yet to be determined. It should be
stressed that now the whole non-Abelian structure of the
dielectric field is taken into account. The electric
part of the gauge potential is equal to zero $A^a_0=0$. Then the
equations of motion (\ref{eqmot1}), (\ref{eqmot2}) take the
following form
\begin{equation}
\left[ \sigma \left( \frac{ h^2 }{\Lambda^2 } \right) g' \right]'
+ \frac{1}{r^2} \sigma \left( \frac{ \phi_a \phi^a }{\Lambda^2 }
\right) g(1-g^2)=h^2g \label{eqmonopol1}
\end{equation}
and
\begin{equation}
-\frac{1}{r^2} (r^2 h')' +\frac{2}{r^4} hg^2 + \frac{1}{2}
\sigma'_h \left[\frac{2g'^2}{r^2} +\frac{(g^2-1)^2}{r^4}
\right]=0. \label{eqmonopol2}
\end{equation}
Here prime denotes differentiation with respect to $r$. The last
equation possesses the obvious solution
\begin{equation}
g=0, \label{solmag1}
\end{equation}
which describes the magnetic monopole located at the origin. It is
the famous Wu--Yang monopole \cite{wu}. Corresponding gauge
potential is singular at the point of the location of the
monopole.
\\
Substituting the obtained solution (\ref{solmag1}) into
(\ref{eqmonopol1}) we obtain the equation for the function $h$
\begin{equation}
-\frac{1}{r^2} (r^2 h')' + \frac{1}{2 r^4} \sigma'_h=0.
\label{eqmonopol3}
\end{equation}
We rewrite it in terms of a new variable $x=\frac{1}{r}$. Then
\begin{equation}
h''_{xx}=\frac{1}{2} \sigma'_{h} . \label{eq_h1}
\end{equation}
It can be easily integrated for any dielectric function $\sigma $
\begin{equation}
h'^2_x = \sigma + C, \label{eq_h2}
\end{equation}
where $C$ is an integration constant. Finally, the solution is
given by the formula
\begin{equation}
\int \frac{dh}{\sqrt{\sigma (h) +D}} = \frac{1}{r \Lambda} +D.
\label{eq_h3}
\end{equation}
Here $D$ is the second integration constant.
\\
In case of the dielectric function (\ref{dick})
the integral (\ref{eq_h3}) can be calculated and we find
\begin{equation}
h=B \left( \frac{1}{r\Lambda} + \frac{1}{\beta_0}
\right)^{\frac{1}{1-4\delta }}, \label{sol_h}
\end{equation}
where $B=|1-4\delta|^{\frac{1}{1-4\delta}}$. To summarize, the
magnetic monopole solution takes the form
\begin{equation}
A^a_i= -\epsilon^{aik} \frac{x^k}{r^2} \label{sol_mag}
\end{equation}
and
\begin{equation}
\phi^a = B \frac{x^a}{r} \left( \frac{1}{r\Lambda} +
\frac{1}{\beta_0} \right)^{\frac{1}{1-4\delta }}.
\label{sol_dick_mag}
\end{equation}
The pertinent energy density reads
\begin{equation}
\varepsilon = B^{8\delta} \frac{1}{r^4} \left( \frac{1}{r\Lambda
}+\frac{1}{\beta_0} \right). \label{en_density_mag}
\end{equation}
It is easy to check that in spite of the singularity in the gauge
potential obtained field configuration has finite total energy for
$\delta > \frac{1}{4} $
\begin{equation}
E_N= \Lambda \frac{4\delta -1}{4\delta +1} B^{8\delta}
\beta_0^{\frac{4\delta +1}{4\delta -1}}. \label{mag_energy}
\end{equation}
Due to the interaction with the dielectric scalar field the
singular Wu--Yang magnetic monopole becomes regularized.
\\
Similarly as in the electric sector there is an infinite energy
solution. It is given by the following scalar field
\begin{equation}
\phi^a (r)= B \Lambda \left( \frac{1}{r\Lambda}
\right)^{\frac{1}{1-4\delta }}. \label{sol_mag_infinit}
\end{equation}
One can notice that these magnetic solutions are BPS
configurations. In order to prove that we will consider the energy
functional in the purely magnetic sector
\begin{equation}
E_N=\int d^3x \left[ \frac{1}{4} \left( \frac{\phi}{\Lambda}
\right)^{8\delta} F_{ij}^aF_{ij}^a +\frac{1}{2} (D_i\phi^a) (D_i
\phi^a) \right]. \label{energy_bogomol1}
\end{equation}
It can be rewritten in the form
\begin{equation}
E_N=\frac{1}{4} \int d^3x  \left[ \left( \frac{\phi}{\Lambda}
\right)^{4\delta} F_{ij}^a -\epsilon_{ijk} (D_k\phi^a) \right]^2
+\frac{1}{2} \int d^3x \left( \frac{\phi}{\Lambda}
\right)^{4\delta} \epsilon_{ijk} F_{ij}^a (D_k\phi^a),
\label{energy_bogomol2}
\end{equation}
then the Bogomolny equation reads
\begin{equation}
\left( \frac{\phi}{\Lambda} \right)^{4\delta} F_{ij}^a =
\epsilon_{ijk} (D_k\phi^a). \label{eq_bogomol1}
\end{equation}
\\
It can be checked by direct calculation that solutions found above
fulfill this condition.
\\
The whole family of our new solutions describing the magnetic
monopole configurations are non-Abelian generalization of the
standard Abelian solution in the Dick model \cite{dick2, dick1}.
As one can easily notice non-Abelian contents of the model does
not change the solutions drastically. They have similar dependence
on the radial coordinate $r$ and in consequence the same
singularities. Due to that one can expect that also the electric
solutions in the full non-Abelian theory will not differ
significantly from their Abelian counterparts. Nonetheless this
prediction has to be checked in an explicit calculation.
\\
It is clearly visible that non-Abelian generalization of the color
dielectric model has little impact on the magnetic monopoles
sector. From this point of view the gain of taking more general
model seems not worth the effort. However, as it will be shown
below, non-Abelian dielectric degrees of freedom modify the
topological contents of the model and play crucial role in the
glueball problem.
\section{ \bf{Faddeev--Niemi action}}
\label{fnsec} Besides the confinement of quarks the glueball
problem i.e. existence of particles build entirely of the gauge
fields is the most striking phenomenon in the non-perturbative
QCD. It is known from lattice theory \cite{lattice-glueball} that
such objects should appear alone or with some non-zero quark
contribution (so-called hybrid states \cite{lattice-hybrid}). On
the other hand, in spite of numerous attempts, the theoretical
understanding of glueballs spectrum, their masses and other
physical features is rather in its infancy.
\\
Between many ideas the proposition to describe glueballs as
topological solitons looks particularly attractive. Such idea has
been used to model hadrons as solitons in the famous Skyrme model
\cite{skyrme}. In case of glueballs it has been suggested that
they should be made of self-linking flux-tubes of the gauge field.
It follows from the observation that in QCD gauge field generated
among a quark and an anti-quark is squeezed into a very thin tube.
Such tubes have to make a loop or other more complicated closed
object since glueballs do not contain quark degrees of freedom.
\\
A model, widely considered in the context of the low energy gluodynamics,
which admits toroidal solitons reads
\begin{equation}
L_{FN}= m^2 (\partial_{\mu } \vec{n} )^2 - \frac{1}{g} [ \vec{n}
\cdot (
\partial_{\mu } \vec{n} \times
\partial_{\nu } \vec{n} )]^2 \label{lag_fn_org}
\end{equation}
and is known as the Faddeev--Niemi model \cite{niemi1, niemi2,
cho}. Here $\vec{n}$ is a real, three component unit vector field,
$m$ is a mass scale and $g$ is a coupling constant. As one can
expect this model possesses non-trivial topology. Static solutions
with $\vec{n} \rightarrow \vec{n}_0 $ for $r \rightarrow \infty$
can be understood as maps from $S^3$ to $S^2$ which are divided
into disconnected homotopy classes $\pi_3(S^2)$ and classified by
the Hopf invariant. In fact, knotted topological solutions in this
model have been recently found \cite{salo, battye}.
\\
It is believed that this model can be obtained by appropriate
decomposition of the gauge fields and integrating out some of the
degrees of freedom. Such decomposition should identify order
parameter which is relevant in the low energy QCD. Its most
commonly accepted gauge covariant form, in case of $SU(2)$ group,
is given by
\begin{equation}
A^a_{\mu }=\epsilon^{abc} n^b \partial_{\mu } n^c +A_{\mu} n^a +
\rho \partial_{\mu } n^a + \sigma \epsilon^{abc} n ^c \partial_{\mu }
n^b  \label{anzatz_fn1},
\end{equation}
where besides previously defined field $\vec{n}$ we have a vector
field $A_{\mu }$ and two scalars $\rho $ and $\sigma $. This
decomposition is motivated by the famous picture of the QCD vacuum
as a condensate of magnetic monopoles (see e.g. \cite{monopole}).
Here, the condensate of monopoles is described by the topological
field $\vec{n}$.
\\
In the context of the non-Abelian color dielectric model one can
generalize the Cho--Faddeev--Niemi Ansatz (\ref{anzatz_fn1}) by
decomposition of the triplet scalar field
\begin{equation}
\phi^a = \phi m^a, \label{azatz_fn2}
\end{equation}
where $\vec{m}$ is a new three component unit vector and $\phi $ a
scalar field. They can be expressed by the primary color dielectric
field in the unique way
$$\phi = \sqrt{\phi_a \phi^a} \; \; \mbox{ and } \; \; m^a
=\frac{\phi^a}{\sqrt{\phi_a\phi^a}}$$ if $\phi^a \neq 0$ for
$a=1,2,3$. In case of vanishing color dielectric field this
decomposition is not well-defined.
\\
Let us now rewrite non-Abelian color dielectric model in terms of
recently introduced variables. The field strength tensor takes the
form
$$
F^a_{\mu \nu } = n^a [F_{\mu \nu} +(1-\rho^2 -\sigma^2)H_{\mu
\nu}] $$
\begin{equation}
+(D_{\mu } \rho \partial_{\nu } n^a -D_{\nu } \rho
\partial_{\mu} n^a) -\epsilon^{abc} n^b (D_{\mu} \sigma \partial_{\nu } n^c -
D_{\nu} \sigma \partial_{\mu } n^c),
 \label{Fdecomp}
\end{equation}
where we have introduced the following abbreviations
$$ F_{\mu \nu }=\partial_{\mu } A_{\nu } -\partial_{\nu }
A_{\mu}, $$
$$ H_{\mu \nu } =\epsilon^{abc} n^a \partial_{\mu } n^b
\partial_{\nu } n^c $$
and
$$ D_{\mu } \rho = \partial_{\mu } \rho +i A_{\mu } \rho.$$
In the same way we express the covariant derivative
$$
D_{\mu } \phi^a = m^a \partial_{\mu } \phi + \phi ( \partial_{\mu
} m^a - m^b n_b \partial_{\mu } n^a + n^a m^b \partial_{\mu } n_b
) - $$
\begin{equation}
-\phi ( \epsilon^{abc} (n^b A_{\mu} + \rho
\partial_{\mu} n^b ) m^c - \sigma m^bn_b \partial_{\mu} n^a +
\sigma n^a m^b \partial_{\mu } n_b ). \label{Ddecomp}
\end{equation}
Finally, after substituting (\ref{Fdecomp}) and (\ref{Ddecomp})
into the Lagrangian (\ref{action}) we get
$$
S= \int d^4 x \sigma \left( \frac{\phi^2}{\Lambda^2} \right)
\left[ n^a [F_{\mu \nu} +(1-\rho^2 -\sigma^2)H_{\mu \nu}] \right.
$$
$$
\left. +(D_{\mu } \rho \partial_{\nu } n^a -D_{\nu } \rho
\partial_{\mu} n^a) -\epsilon^{abc} n^b (D_{\mu} \sigma \partial_{\nu } n^c -
D_{\nu} \sigma \partial_{\mu } n^c)] \right]^2 + $$
$$
\left[ m^a \partial_{\mu } \phi + \phi ( \partial_{\mu } m^a - m^b
n_b \partial_{\mu } n^a + n^a m^b \partial_{\mu } n_b ) - \right.
$$
\begin{equation}
\left. - \phi ( \epsilon^{abc} (n^b A_{\mu} + \rho
\partial_{\mu} n^b ) m^c - \sigma m^bn_b \partial_{\mu} n^a +
\sigma n^a m^b \partial_{\mu } n_b ) \right]^2 \label{Ldecomp}
\end{equation}
Now, we get rid of some degrees of freedom which are supposed to
play marginal role in the low energy limit i.e. we put $A_{\mu} =
\rho = \sigma =0$. In the other words, we construct a constrained
model where only topological field $\vec{n}$ is left. Moreover,
we assume that the scalar field $\phi $ condensates in its vacuum
on a constant, non-zero value $\phi_0$
\begin{equation}
\phi = \phi_0 = \mbox{const.} \label{phi_limit}
\end{equation}
Then the model takes the form
$$
\mathcal{L}= -\frac{\sigma_0  }{4} [ \vec{n} \cdot ( \partial_{\mu
} \vec{n} \times \partial_{\nu } \vec{n} )]^2 + (\partial_{\mu }
\vec{n} )^2 (\vec{n} \cdot \vec{m} )^2 \phi_0^2 + \phi_0^2
(\partial_{\mu } \vec{n} \cdot \vec{m} )^2 -$$
\begin{equation}
-2\phi_0^2 \vec{n} \cdot \vec{m} (\partial_{\mu } \vec{n} \cdot
\partial^{\mu } \vec{m} ) +2\phi_0^2 (\partial_{\mu } \vec{n} \cdot
\vec{m} ) (
\partial^{\mu } \vec{m} \cdot \vec{n} ) + \phi_0^2 (\partial_{\mu
} \vec{m} )^2, \label{lag_fn1}
\end{equation}
where $\sigma_0=\sigma ( \phi_0 )$. The last step to derive
the Faddeev--Niemi model is to assume that vector field $\vec{m}$
condenses as well
\begin{equation}
\vec{m} = \vec{m}_{\infty} = \mbox{const.}  \label{m_limit}
\end{equation}
It is equivalent to the condensation of all components of the
primary color dielectric field $\phi^a$. Eventually we find
\begin{equation}
\mathcal{L}=-\frac{\sigma_0  }{4} [ \vec{n} \cdot ( \partial_{\mu
} \vec{n} \times \partial_{\nu } \vec{n} )]^2 + (\partial_{\mu }
\vec{n} )^2 (\vec{n} \cdot \vec{m}_{\infty} )^2 \phi_0^2 +
\phi_0^2 (\partial_{\mu } \vec{n} \cdot \vec{m}_{\infty} )^2.
\label{lag_fn2}
\end{equation}
Here, one should make a few remarks concerning the Lagrangian
obtained above.
\\
First of all, it is expected that in the original Faddeev--Niemi
model the appearance of the dimensional parameter $m$ (i.e.
existence of the usual kinetic term for the unit vector field) is
due to integrating out the Abelian Higgs multiplet $(A_{\mu},
\rho, \sigma)$ from the full quantum theory. On the contrary, in
the non-Abelian color dielectric model it is sufficient just to
neglect these degrees of freedom. However, this 'contradiction',
can be easily explained. As we have noticed before, the
non-Abelian color dielectric model is described by a classical
effective action which has been postulated using some arguments
from lattice field theory. The construction of the effective model
was based on blocking procedure of the gauge fields on the
lattice. In our framework the blocking plays identical role as
integrating out some quantum fields. Thus, to find action for the
unit vector field we should neglect non-topological fields -- not
to integrate them. In some sense, it was already done. Because of
that, the problem known from the Faddeev--Niemi model concerning
correct integration of Abelian Higgs multiplet can be formulated
here as a problem of deriving the non-Abelian color dielectric
Lagrangian from QCD. Due to the fact that the mass parameter is
given by the vacuum value of the dielectric field $\vec{\phi}$ a
particular form of the potential $V(\vec{\phi} )$ is needed.
\\
Secondly, one can notice the fundamental difference between the
original Faddeev--Niemi model and our proposal. Namely, the
Faddeev--Niemi action is invariant under $O(3)$ rotations. On
account of the fact that the invariance group of the ground state
is  $O(2)$ the spontaneous symmetry breaking occurs and two
Goldstone bosons should emerge. Moreover, there is no mass gap in
the spectrum of excitation of this model. Such problems can
disappear in the model postulated here because of the condensation
of the color dielectric field $\vec{m}$. Our model breaks $O(3)$
symmetry explicitly on the Lagrangian level and one can expect
that no Goldstone boson appears. In fact, it has been confirmed by
Dittmann et. al. in a similar model \cite{wipf} (they included
symmetry breaking terms with a source filed $\vec{h}$ in the
Faddeev--Niemi action, which in our model is just the condensation
value of the field $\vec{m}$). In addition, they observed a mass
gap as well.
\\
Let us notice that the symmetry breaking is due to the very
non-trivial, dielectric-like term. It is unlikely the standard
procedure where the symmetry breaking was given by some potential
terms \cite{niemi3}, \cite{sanchez}. One should remember that the
breaking of the symmetry and removing of the Goldstone bosons is
not sufficient to exclude all massless excitations. There is still
a chance to have such solutions. Due to that the existence of the
mass gap is still an open problem.

\section{\bf{Conclusions}}
In the paper the minimal non-Abelian generalization of the color
dielectric model has been proposed. Using some arguments from the
lattice gauge theory we argue that the model should consist of
$SU(2)$ gauge field and non-Abelian color dielectric field. On
account of the fact that this dielectric field belongs to the
fundamental representation of the $SU(2)$ group its kinetic term
is given by the covariant derivative instead of the standard one.
That makes the coupling between gauge and color dielectric fields
double folded -- minimally by the covariant derivative and
non-minimally by color dielectric function. This last coupling is
assumed to be identical as in the standard Abelian color
dielectric case.
\\
It has been shown that such model can serve well to reproduce
confinement of external electric sources already on the classical
level. Discussion of the electric sector has been carried out in
the Abelian sector of our model. Even in this restricted theory
there is a pure electric configuration generated by the external
static point-like source having infinite total energy. However, in
contradiction to the usual Maxwell electrodynamics, the
singularity appears due to the long range behavior of the fields.
We have found that electric potential (and energy  in the vicinity
of  the charge) grows as $r^{\alpha}$, where $\alpha \in (0,1)$,
for the dielectric function (\ref{dick}). That is in a good
agreement with phenomenological data and the latest theoretical
considerations. In addition, there exists single-parameter family
of finite energy solutions. Analogously, finite as well as
infinite energy solution has been found in the magnetic sector of
the theory. In this case restriction to the Abelian sector is no
longer needed and one can find magnetic monopole solution
surrounded by the non-Abelian color dielectric field. We have also
proved that they are BPS solution fulfilling the corresponding
Bogomolny equations. It is easy to notice that adding a potential
term to the action will obviously fix the asymptotic value of the
dielectric field and in consequence, from the whole family of
solutions only one will be preserved.
\\
We have also observed that the proposed model gives, in the limit
when the color dielectric field condensates and gauge field is
constrained only to the so-called topological degrees of freedom,
a modified Faddeev--Niemi Lagrangian which possesses toroidal
soliton solutions interpreted as glueballs. Our modification
breaks $O(3)$ symmetry explicitly on the Lagrangian level. This is
a great advantage of the model since the massless Goldstone bosons
are excluded.
\\
To summarize, the non-Abelian color dielectric model seems to be a
pretty good candidate for the correct effective model for the low
energy gluodynamics. It describes simultaneously quark confinement
(with potential consistent with experimental data) and glueball
states. To the best of our knowledge this is the first model which
is able to joint these features. It clearly exposes the necessity
of taking into account the full set of non-Abelian degrees of
freedom in the framework of the color dielectric approach. Even in
the first, naive attempt such a theory is considerably better
suited to description of non-perturbative gluonic dynamics than
commonly used Abelian color dielectric models. Therefore it shows
the direction in which the progress can be achieved in the future.
\\
Of course, there are a lot of questions which still need to be
answered. First of all, one has to get rid of the finite energy
solutions (electric and magnetic). It can be done by inserting a
potential term into the Lagrangian which would force vanishing of
the scalar field at the spatial infinity. On the other hand one
can observe that this makes the glueball sector trivial. In our
approach the glueball spectrum is strongly dependent on the vacuum
value of the scalar field. It is possible, for zero vacuum value
of the scalar fields, to trivialize any knot soliton -- it costs
zero energy to untie any object of this kind. It could be cured by
a more complicated potential with two minima -- for zero and
non-zero scalar field. Then confining and glueball solutions would
appear in two different phases. It does not seem to be a
satisfactory solution to this issue. We believe that more subtle
mechanism can be responsible for making finite energy solutions
unstable and for removing them from the physical spectrum of the
theory. Moreover, one can take advantage of the approach recently
proposed by Bazeia and collaborators and analyze dynamical sources
(quarks) \cite{bazeia2}.
\\
Secondly, the influence of the Abelian Higgs multiplet on the
glueball sector should be studied. In particular, one has to
clarified the role of the Abelian gauge field $A_{\mu}$. Presence
of this field is crucial for preserving  gauge invariance after
performing the decomposition.
\\
The Faddeev--Niemi model with explicitly broken $O(3)$ symmetry
also needs more detailed studies.
\\
However, in our opinion the most urgent challenge in the presented
approach is to get deeper insight into the relation of our
effective model and the underlying quantum theory. We plan to
address this issue in the nearest future.
\\
To conclude, the model considered in our paper should be treated
as a first but quite encouraging step on the way to the correct
effective theory for the low energy gluodynamics. Further
exploration of this area is undoubtedly mandatory.
\\
\emph{Acknowledgements:} A.~W. gratefully acknowledges the support
from the Foundation for Polish Science.

\end{document}